\documentclass[twocolumn,prl,amsmath,amssymb,showkeys,showpacs]{revtex4}

\usepackage{graphicx}

\usepackage{dcolumn}

\usepackage{bm}

\begin{document}

\title{Observing Quantum Vacuum Lensing in a Neutron Star Binary System}

\author{Arnaud Dupays}
\affiliation{
Laboratoire Collisions, Agr\'egats, R\'eactivit\'e, IRSAMC,
CNRS/UPS, 31062 Toulouse, France
}
\author{C\'ecile Robilliard}
\affiliation{
Laboratoire Collisions, Agr\'egats, R\'eactivit\'e, IRSAMC,
CNRS/UPS, 31062 Toulouse, France
}
\author{Carlo Rizzo}
\affiliation{
Laboratoire Collisions, Agr\'egats, R\'eactivit\'e, IRSAMC,
CNRS/UPS, 31062 Toulouse, France
}
\author{Giovanni F. Bignami}
\affiliation{
Centre d'Etude Spatiale des Rayonnements,
CNRS/UPS, 31401 Toulouse, France \\
Dipartimento di Fisica Nucleare e Teorica, Universita di Pavia, Italy
}

\date{\today}

\begin{abstract}
In this letter we study the propagation of light in the
neighbourhood of magnetised neutron stars. Thanks to the optical
properties of quantum vacuum in the presence of a magnetic field,
light emitted by background astronomical objects is deviated,
giving rise to a phenomenon of the same kind as the gravitational
one. We give a quantitative estimation of this effect and we
discuss the possibility of its observation. We show that this
effect could be detected by monitoring the evolution of the recently
discovered double neutron star system J0737-3039.
\end{abstract}

\pacs{}

\maketitle

Neutron stars have been intensively studied since their discovery
as pulsars, i.e. stars with a magnetic dipole field tilted with
respect to their rotation axis, emitting radio waves along their 
dipole axis as the star spins \cite{ReviewPulsar}. Typically their
mass is on the order of that of the Sun, and their radius of the
order of $10$ km. Their spinning period ranges from tens of
milliseconds for young stars to seconds, decreasing with star age.
The magnetic field of neutron stars can be as high as $10^{11}$ T
as in the case of magnetars \cite{IbrahimBignami} while magnetic
fields of the order of $10^8-10^9$ T are more common. Most of the
known pulsars are isolated stars, less than 100 are known to be in
binary systems with non-degenerate stars, few ($\sim 5$) are
in neutron star-neutron star binary systems. Very recently, a
specially interesting NS-NS system (J0737-3039) has been
discovered \cite{DoublePulsar}.

On the other hand, it is known that quantum vacuum is a non
linear optical medium \cite{Euler}. In the seventies it was
clearly pointed out \cite{Bialynicka} that the velocity of light
propagating in a quantum vacuum in the presence of a magnetic
field changes of a quantity which depends on the polarization of
light with respect to the magnetic field direction. Those
calculations were prompted by early speculations that neutron
stars could be a source of very intense magnetic fields. Since
then, most authors have been interested in how the magnetic
field can affect the propagation of radiation emitted by the
neutron star itself \cite{emission}.

In this letter we investigate the case of radiation emitted by
another astronomical object passing close enough to the NS to be
significantly  affected by its magnetic field. The case in point 
will be seen to be that of J0737, especially in
the context of its orbital evolution.

In the vicinity of a NS, owing to the optical properties of
quantum vacuum in the presence of a magnetic field, light emitted
by astronomical sources is deviated, giving rise to a phenomenon of
the same kind as the gravitational one predicted by general
relativity. We first prove that fields of the order of $10^9$ T are
not sufficient to give an effect big enough to be detected in the
case of isolated NS or in the case of NS-non-degenerate star
systems. On the other hand we show that quantum vacuum effect could be
observed by monitoring the evolution of the recently discovered NS-NS 
system J0737-3039.

It is known that light generally bends towards regions with a
higher index of refraction \cite{BornWolf}. Because of refraction
index variation in magnetized quantum vacuum, light passing 
near a magnetar behaves as if it were attracted by the magnetar and it should bend
considerably, giving rise to a lensing effect. Actually, since the
velocity of light depends on its polarization, the cosmos around a
magnetar acts as a birefringent medium and different polarizations
should bend differently.

In the presence of a static magnetic field $\mathbf{B}$ the index
of refraction for light polarized parallel (resp. perpendicular)
to the magnetic field $n_{\|(\bot)}$ can be written as
\cite{Bialynicka}

\begin{equation}
n_{\|(\bot)} =   1 + a_{\|(\bot)} {B}^2
\end{equation}

where $a_{\|}$ is about $9 \times 10^{-24}$ T$^{-2}$ and
$a_{\bot}$ is about $5 \times 10^{-24}$ T$^{-2}$. Deviations
from these formulae appear for magnetic field values bigger than
the critical field  $4.4\, 10^{9}$ T \cite{Heyl} when a slower
  dependence on $B$ sets in. However, in our case,
  since we are interested in subcritical fields, such deviation is negligible.
This variation of the index of refraction of light in vacuum is
very small when the magnetic field applied is the order of the
fields that can be produced in a terrestrial laboratory (about 10
T), so that this effect has not yet been detected \cite{expVMB}.
For our purpose, this effect can be considered achromatic because this formula
applies up to photon energies much larger than optical energies
\cite{Tsai}.

For sake of simplicity, in what follows we will neglect intrinsic 
polarization of light, considering that astronomical sources are weakly
polarized.

Propagation of light can be studied using the differential
equation of light rays \cite{BornWolf}. The specific case of light
propagating in a plane perpendicular to the NS magnetic dipole moment
has been treated in ref. \cite{ArticoloRussi}. In
ref. \cite{ArticoloRussi} it was concluded that a neutron star with a magnetic field of around
$10^9$ T could give a detectable effect.

\begin{figure}[h]
\includegraphics[width=8cm]{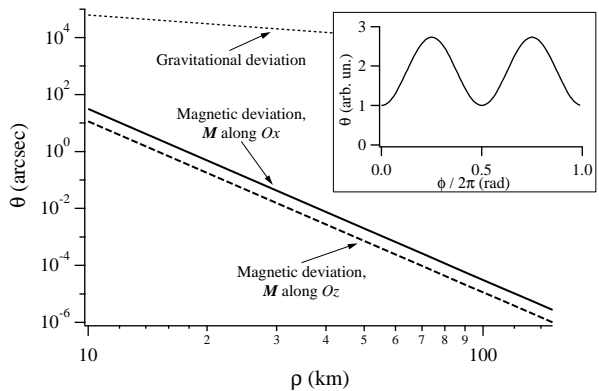}
\caption{\label{deviation}{Deviation angle of a light ray
propagating parallel to the Oz axis with a minimal distance $\rho$
to the center of the NS, for two different orientations of the
magnetic dipole moment $M$. The value of the magnetic field at the
surface of the star is $B_0=10^{9}$ T. Inset shows deviation
modulation with NS rotational phase.}}
\end{figure}

In Fig. \ref{deviation}, we present the results of our
calculations of the deviation angle of light ray for different
orientations of the magnetic dipole moment, in the case of a
magnetic field of $10^9$ T. This shows that the deviation angle
goes as $1/\rho^6$, where $\rho$ is the minimal distance
between the light ray and the neutron star.

This figure also shows that the deviation angle depends on the
orientation of the magnetic moment. Since the neutron star spins,
the light deviation varies in time at twice the frequency of
the spinning pulsar, owing to the spatial symmetry of the dipole
field. Depending on orientation, the modulation of the effect can be
as high as $40\%$. Magnetic deviations at very short
distances from the neutron star become of the same order as the
gravitational one only in the case of magnetar ($B_0=10^{11}$ T), i.e
$\theta \sim 10^5$ arcsec.
In the more common cases of a field $B_0=10^{8}-10^{9}$ T, the magnetic
deviation is always very small compared to the gravitational one (see
Fig. \ref{deviation}).
On the other hand, for astronomical distances, even very small
deflection can completely change observations. It is also
important to note that magnetic birefringence exists in any
medium (Cotton-Mouton effect), and in particular in gases. The
residual gas that fills the space around neutron stars is also
magnetically birefringent. The density of this gas is between 1
and $10^4$ particles per $cm^3$. Thanks to the value of the
Cotton-Mouton constant for simple gases like hydrogen
\cite{RizzoRizzoBishop}, the effect induced by this quantity of
gas will be at least 6 orders of magnitude smaller that the one
induced by the quantum vacuum. This effect will thus be neglected here.

\begin{figure}[h]
\includegraphics[width=8cm]{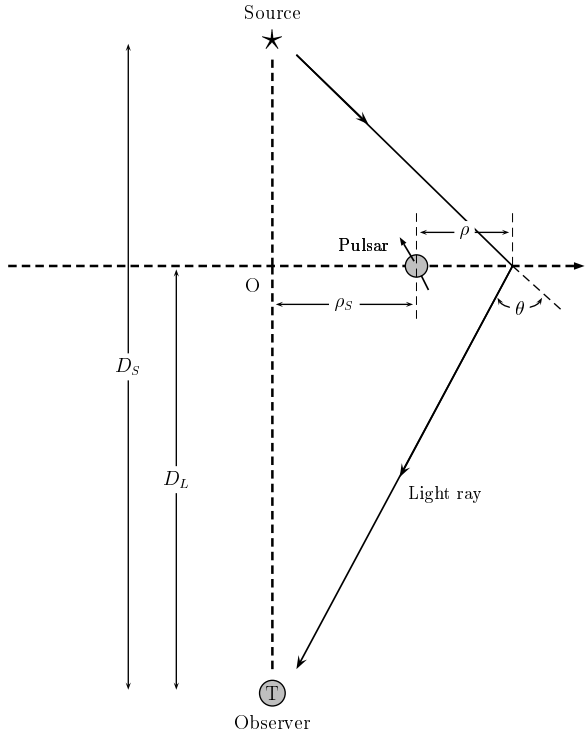}
\caption{\label{schema}{Geometry of the deflection effect.
The light rays are deflected near the pulsar by the angle $\theta$.}}
\end{figure}

To study the feasibility of the observation of quantum vacuum
mirage, we started with the case of an isolated neutron star with a
distant background source (see Fig. \ref{schema}). To estimate the quantum
vacuum lensing, we have performed numerical calculations. Following
the method proposed by Paczy\'{n}ski \cite{Pacz:96} for gravitational 
lensing, we have included the quantum vacuum effect
in the total angular deviation of light rays. The problem consists
of calculating the imaging of a distant source to detect
modifications to the magnification due to gravitational
lensing. For sake of simplicity we have considered the case of a
magnetic dipole moment parallel to the light propagation direction. 
The deflection is then given by :
\begin{equation}
\label{eqdev}
\theta=\frac{4GM}{\rho c^2}+\frac{5\pi a B_0^2\rho_0^6}{\rho^6}
\end{equation}
where $G$ is the gravitational constant, $M$ and $\rho_0$ the mass
and the radius of the pulsar respectively, and $B_0$ the value of
the magnetic field at the surface of the star. In our simulations
$a$ has been taken equal to $a_{\bot}$. Finally, $\rho$ is the distance
from the deflected light ray to the center of the pulsar.
Fig. \ref{schema} presents the geometry of the system, projected on the
trajectory plane of the light ray. The distances from the observer
to the lens (pulsar) and to the source are indicated as $D_L$ and $D_S$,
respectively. $\rho_S$ is the distance from the pulsar to the line of sight.

Defining $D=(D_L-D_S)D_L/D_S$, the linear Einstein ring radius is
then given by $R_E=\sqrt{4GMD/c^2}$. For a distance from the
pulsar to the line of sight $\rho_S$ larger than $R_E$, the
gravitational lensing is very small \cite{Pacz:96}. With $M\sim
M_\odot$, $\rho_0\sim 10$ km and $B_0=10^{8}-10^9$ T (which corresponds
to a reasonable value of the magnetic field at the surface of a
neutron star), the calculations show that $\rho_S \gg R_E$ : the
magnification is then negligible for both gravitational and total
lensing. The next case to be studied was a binary system consisting
of a neutron star with a normal star companion. With respect to
Fig. \ref{schema}, $D$ changes with the orbital trajectory, but
the result is similar to the previous case. To be able to observe
any modification of the gravitational lensing due to magnetic
effects of the order of few percent, the latter have to occur at
distances of the order of $R_E$. These conditions begin to be
realized for a totally non realistic magnetic field $B_0\sim 10^{16}$ T. The
effect of $10^8$ T corresponds to negligible contributions. 
Our results are here in disagreement with the conclusions of
authors of ref. \cite{ArticoloRussi}. In any case, in a realistic astronomical case, the effects of
accretion onto the NS of mass transferred from the companion will
completely obliterate subtle effects of the type we are interested
in. 

Much more interesing appears to be the case of the NS-NS
binary J0737. No accretion perturbs this system which, furthermore,
presents a very favourable orbital inclination (currently, $i \sim
87^{\circ}$) \cite{NSNSgeometry}. In fact, orbital precession of the
system allows one to predict that, before 2020, $i$ will be
extremely close to $90^{\circ}$. The two NS will actually
geometrically "eclipse" each other, as seen by the earth, for a
short time. In any case they will remain practically aligned for
several revolutions of their 2 hours and 45 minutes orbit during a
time long enough to allow for meaningful observations.

Nowadays the radio beam, emitted by neutron star A, passes
at around 20000 km from the surface of neutron star B. The beam is attenuated
or deviated in such a way that eclipses are visible from the
earth. Recently, a modulation of the beam at twice the spinning
frequency of the neutron star B has been observed
\cite{ModDoublePulsar}. The explanation of this effect is the
interaction between light and the magnetosphere of star B. We also know
that this absorption should disappear for photons above about 7.5 GHz
\cite{GreenBank}. The main difference between this case and the
previous two is that the radiation emitted by neutron star A
can be treated as a single directional light ray. We can thus
directly observe the lensing effect due to gravitational and
magnetic deflections of light by pulsar B.

\begin{figure}[h]
\includegraphics[width=8cm]{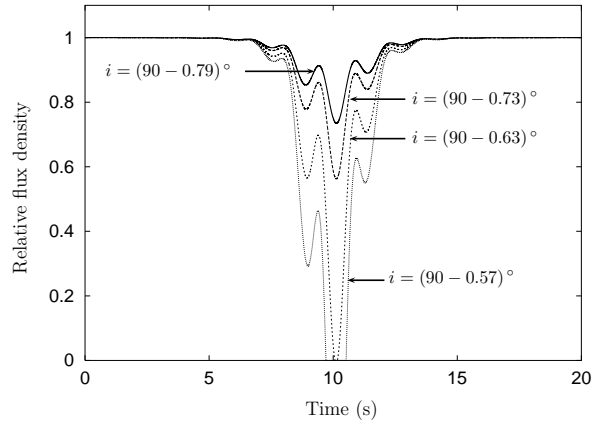}
\caption{\label{fignsns1}{Relative flux density of pulsar A versus time for
    different values of the orbital inclination $i$. The value of the
    magnetic field at the surface of pulsar B is $B_0=10^8$T. Flux densities have
    been normalized to unity when the magnetic effect is negligible. Gravitational lensing is taken into account.}}
\end{figure}

\begin{figure}[h]
\includegraphics[width=8cm]{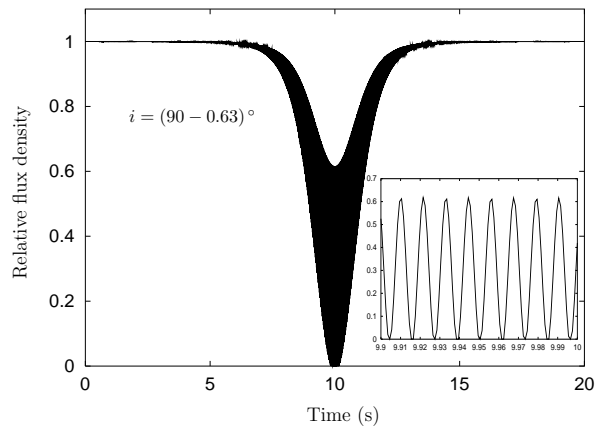}
\caption{\label{fignsns2}{Relative flux density of pulsar B versus
    time for an orbital inclination $i=(90-0.63)^{\circ}$. The value of the
    magnetic field at the surface of pulsar A is $B_0=10^8$T. Flux densities have
    been normalized to unity when the magnetic effect is
    negligible. Gravitational lensing is taken into account. Inset shows the small modulation period of the signal
    ($\sim 11.5$ ms). This modulation is so rapid that it appears as a
    continuum in the main figure.}}
\end{figure}

In Fig. \ref{fignsns1} and Fig. \ref{fignsns2} we present the results
of our calculations. The relative flux density of pulsar A arriving on Earth for
different values of $i$ versus time is shown Fig. \ref{fignsns1}. The calculations assume photons
to be energetic enough not to interact with the magnetosphere of
pulsar B. 
The results have been obtained using the
geometry data published in \cite{NSNSgeometry}. The modulation
shown by the signal is due to the spin period of 2.77 s of pulsar
B. Since the spin period of pulsar A is smaller than the one of pulsar
B ($23$ ms), pulsar A will produce a similar effect on the flux
density of pulsar B, but with the smaller modulation period (see
Fig. \ref{fignsns2}). Total eclipse occurs when the bending is so
large that the beam no longer intersect the line of sight.
Thus, quantum vacuum lensing will give an effect similar
to the radio eclipses observed by \cite{ModDoublePulsar}, of course on photons which should
not be eclipsed. Thanks to the fact that quantum vacuum lensing
is achromatic, this also applies to photons of higher energy, such as
X-rays, also observed from this system \cite{Pell:04}. 
Any observation, once $B_0$ is known, will
provide the value of $a$, which has never been measured. A measure of
$a$ could confirm the classical QED predictions given in
\cite{Euler}. However, we note that magnetic birefringence could also
be induced by physics beyond the standard model, as in the case of the
existence of axions, as shown in
\cite{Maiani}.

In conclusion, we have shown that the magneto-optical properties
of the quantum vacuum induced by the very high magnetic field in
the vicinity of a neutron star can affect the propagation of light
passing near such a pulsar. It will give rise to a lensing effect that
could be detected on earth by monitoring the double neutron star
system J0737-3039, recently discovered. Its observation would
consitute the first experimental confirmation of quantum vacuum optical
properties. Thus, a continuous monitoring of this very peculiar
system will be valuable to test quantum electrodynamic theory, and beyond.




\end{document}